\documentstyle[epsf,12pt]{article}
\begin{document}
\begin{center}
\Large{\bf How to Measure Squeeze Out.}\\
\large{R.S. Longacre$^a$\\
$^a$Brookhaven National Laboratory, Upton, NY 11973, USA}
\end{center}
 
\begin{abstract}
Squeeze out happen when the expanding central fireball flows around a large
surface flux tube in a central Au-Au collision at RHIC. We model such an 
effect in a flux tube model. Two particle correlations with respect to the
$v_2$ axis formed by the soft fireball particles flowing around this large
flux tube is a way of measuring the effect.
\end{abstract}
 
\section{Introduction} 

The flux tube model does well in describing a central Au-Au collision at 
RHIC\cite{QGP,tubevsjet}. However the tubes on the inside of 
colliding central Au-Au will undergo plasma instabilities\cite{PI1,PI2} and
create a locally thermalized system. A hydro system with transverse flow builds
causing a radially flowing blast wave\cite{blast}. The flux tubes that are near
the surface of the fireball gets the largest radial flow and are emitting 
particles from the surface. The hydro flow of particles quarks and gluons
around a larger flux tube on the fireball surface we will call squeeze out.
This is an analogy to tooth paste squeezing out the sides of a tooth paste 
tube opening when the opening is blocked by a dried and harden tooth paste
plug. This squeeze out has been simulated in Ref.\cite{shadowing} but was 
called flux tube shadowing. The major conclusion of flux tube shadowing from
the above reference is the development of a strong $v_3$ azimuthal 
correlation. This $v_3$ will extend over the length of the flux tube which was
created by the initial glasma conditions\cite{QGP}. The flux tube model of
Ref.\cite{tubevsjet} assumed that the locally thermalized system formed 
spherically symmetric blast wave. We need to add to this model a hydro squeeze 
out around the largest flux tube. By hand we will add this modification and
explore possible two particle correlations with respect the flow axis($v_2$)
defined by the squeeze out particles. This flow axis will be at right angles
the the largest flux tube.

The paper is organized in the following manner:

Sec. 1 is the introduction of the squeeze out. Sec. 2 presents the flux tube
model and how squeeze out is added. Sec. 3 defines an angular correlation 
between $v_3$ and $p_t$ dependent $v_1$ with respect to the $v_2$ reaction 
plane Sec. 4 defines an angular correlation between $v_1$ and $p_t$ dependent 
$v_1$ with respect to the $v_2$ reaction plane.  Sec. 5 explores two major 
components to the angular correlation between $v_3$ and $p_t$ dependent $v_1$ 
with respect to the $v_2$ reaction plane. Sec. 6 explores the same two major 
components to the angular correlation between $v_1$ and $p_t$ dependent $v_1$ 
with respect to the $v_2$ reaction plane. Sec. 7 presents the summary and 
discussion.

\section{Flux Tube Model and the addition of Squeeze Out} 
The Glasma flux tube model for RHIC\cite{QGP,tubevsjet} central Au-Au collision
at $\sqrt{s_{NN}}$ = 200.0 GeV has an arrangement of surface flux tubes
which are formed in the initial collision and conserve momentum between them.
The tubes expand longitudinally and are pushed out radially. The higher $p_t$
particles come mainly from particle emission from the tubes. Some arrangements
of the tubes favor a large $v_2$ among the higher $p_t$ particles while other 
arrangements of the tubes favor a large $v_3$(see the top two figures of 
Figure 1.). 

The Flux Tube Model of Ref.\cite{tubevsjet} approximate the soft particle of 
the hydro system with particle production given by Ref.\cite{hijing}. In each 
central event we rotate the largest flux tube into the X axis as shown in 
Figure 1. Each particle from the flux tube fragmentation is given a unique tag,
where the largest flux tube has tag number 1. 

Next we adjust the momentum vector direction of some of the soft particles 
such that we achieve the flow pattern of the lower figure of Figure 1. We 
use a particle by particle one up the Y axis and then one down the Y axis.
One flowing towards +$45^0$ and one flowing towards -$45^0$. A $180^0$ 
particle of around twice the momentum is moved into place to conserve momentum.
Each of the $\pm45^0$ and $180^0$ is given a tag. The flux tubes has a built 
in momentum conservation however we must make a final adjustment of the soft 
particles in order to conserve momentum in the X and Y direction. This is done 
by simply scaling up or down the positive components of the soft particles 
This is a very small change because of the care used in imposing the squeeze 
out flow. We sum over many generated central event to make sure that the 
overall $v_2$ lies along the Y axis, while the $v_3$ is pointed in the $180^0$ 
direction or the minus X axis.  

\begin{figure}
\begin{center}
\mbox{
   \epsfysize 8.0in
   \epsfbox{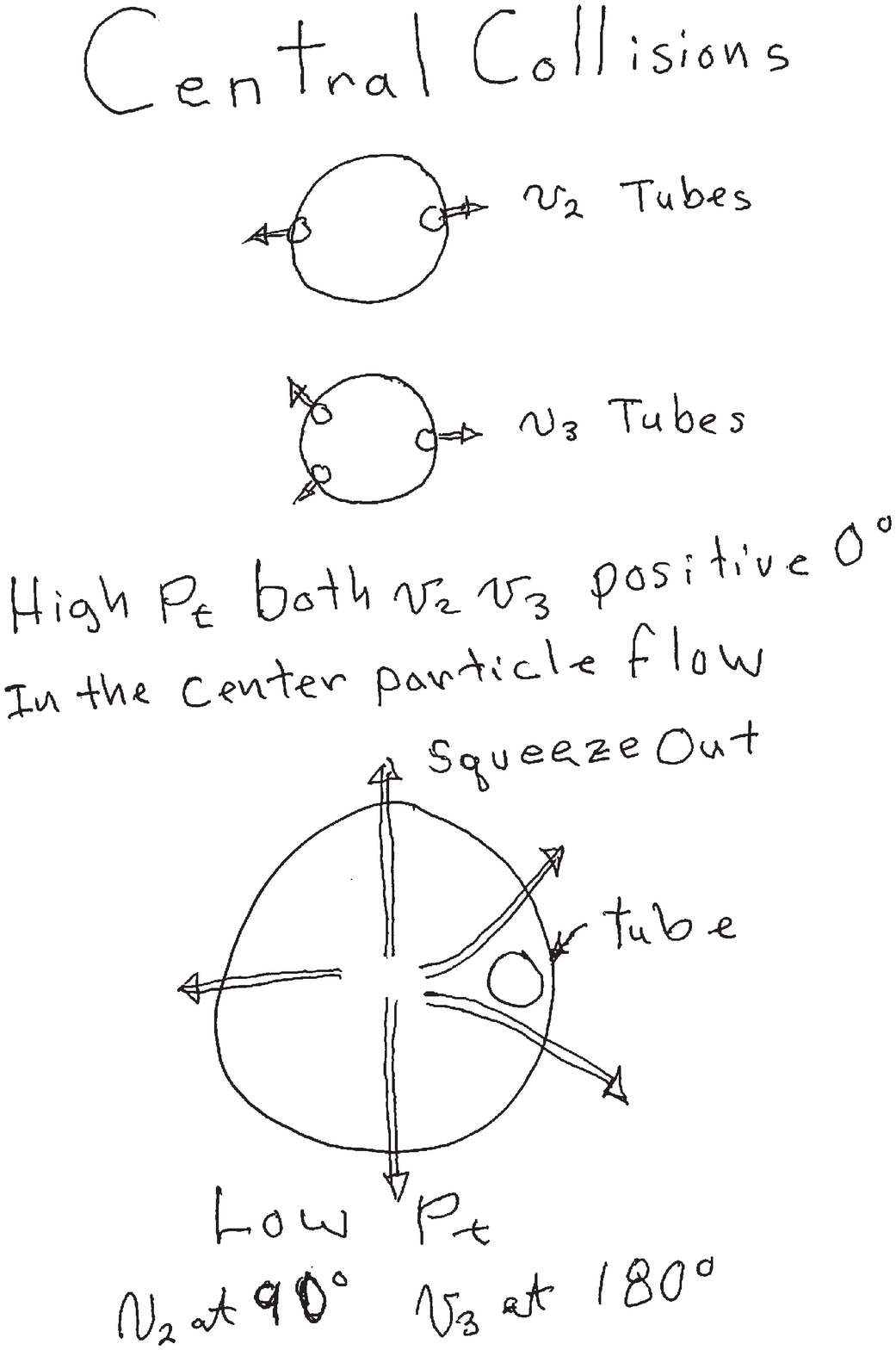}}
\end{center}
\vspace{2pt}
\caption{ Flux Tubes on the surface are arranged around the boundary with some 
events having a strong $v_2$, while other events may have a stronger $v_3$ 
with respect to the largest Tube(see top two figures). This makes a $v_2$ and 
a $v_3$ aligned at $0^0$ at high $p_t$. Soft particles flowing around the 
largest Flux Tube creates at low $p_t$ a $v_2$ at $90^0$ to the largest Tube 
and a $v_3$ at $180^0$(see bottom figure).}
\label{fig1}
\end{figure}

\section{Angular Correlation between $v_3$ and $p_t$ dependent $v_1$ with
respect to the $v_2$ reaction plane}

In the last section we created our simulation of squeeze out to have a global 
$v_3$ pointed in the minus X direction(see lower figure of Figure 1.). At
low $p_t$ $v_1$ will also point in this minus X direction. However as we move
to higher $p_t$ the largest flux tube will take over and $v_1$ will switch to
the positive X axis. We can capture this behavior through a two particle
correlation  with respect to the reaction plane($\Psi_{RP}$) given by global 
$v_2$.

\begin{equation}
\langle cos(3\phi_1 - \phi_2(p_t) - 2\Psi_{RP})\rangle ,
\end{equation}
where $\Psi_{RP}$, $\phi_1$, $\phi_2$ denote the azimuthal angles of the
reaction plane, produced particle 1, and produced particle 2. This two
particle azimuthal correlation measures the difference between the global $v_3$
and the $p_t$ dependent $v_1$. If we would rotate all events such that 
$\Psi_{RP}$ = 0.0, then we have
\begin{equation}
\langle cos(3\phi_1 - \phi_2(p_t))\rangle.
\end{equation}

We calculate this correlation over the simulated central events and obtain the
values shown in Figure 2. where the $p_t$ is that of the second particle.
This sign change moving from lower $p_t$ to higher $p_t$ is the shift of $v_1$
from aligned with global $v_3$ at lower $p_t$ to anti-aligned at higher $p_t$.
more insight on this will be seen in later sections.
 
\section{Angular Correlation between $v_1$ and $p_t$ dependent $v_1$ with
respect to the $v_2$ reaction plane}

At low $p_t$ $v_1$ will points in this minus X direction. However as we move
to higher $p_t$ the largest flux tube will take over and $v_1$ will switch to
the positive X axis. However the global $v_1$ of the event is not so simple to
understand. Let us form a two particle correlation with respect to the 
reaction plane($\Psi_{RP}$) given by global $v_2$, and take the sum of global
$v_1$ vs $p_t$ dependent $v_1$,

\begin{equation}
\langle cos(\phi_1 + \phi_2(p_t) - 2\Psi_{RP})\rangle ,
\end{equation}
where $\Psi_{RP}$, $\phi_1$, $\phi_2$ denote the azimuthal angles of the
reaction plane, produced particle 1, and produced particle 2. This two
particle azimuthal correlation measures the sum between the global $v_1$
and the $p_t$ dependent $v_1$. If we would rotate all events such that 
$\Psi_{RP}$ = 0.0, then we have
\begin{equation}
\langle cos(\phi_1 + \phi_2(p_t))\rangle.
\end{equation}

We calculate this correlation over the simulated central events and obtain the
values shown in Figure 3. where the $p_t$ is that of the second particle.
This sign does not change moving from lower $p_t$ to higher $p_t$, however 
there is a shift toward the positive. More insight on this will be seen in 
later sections.

\begin{figure}
\begin{center}
\mbox{
   \epsfysize 7.0in
   \epsfbox{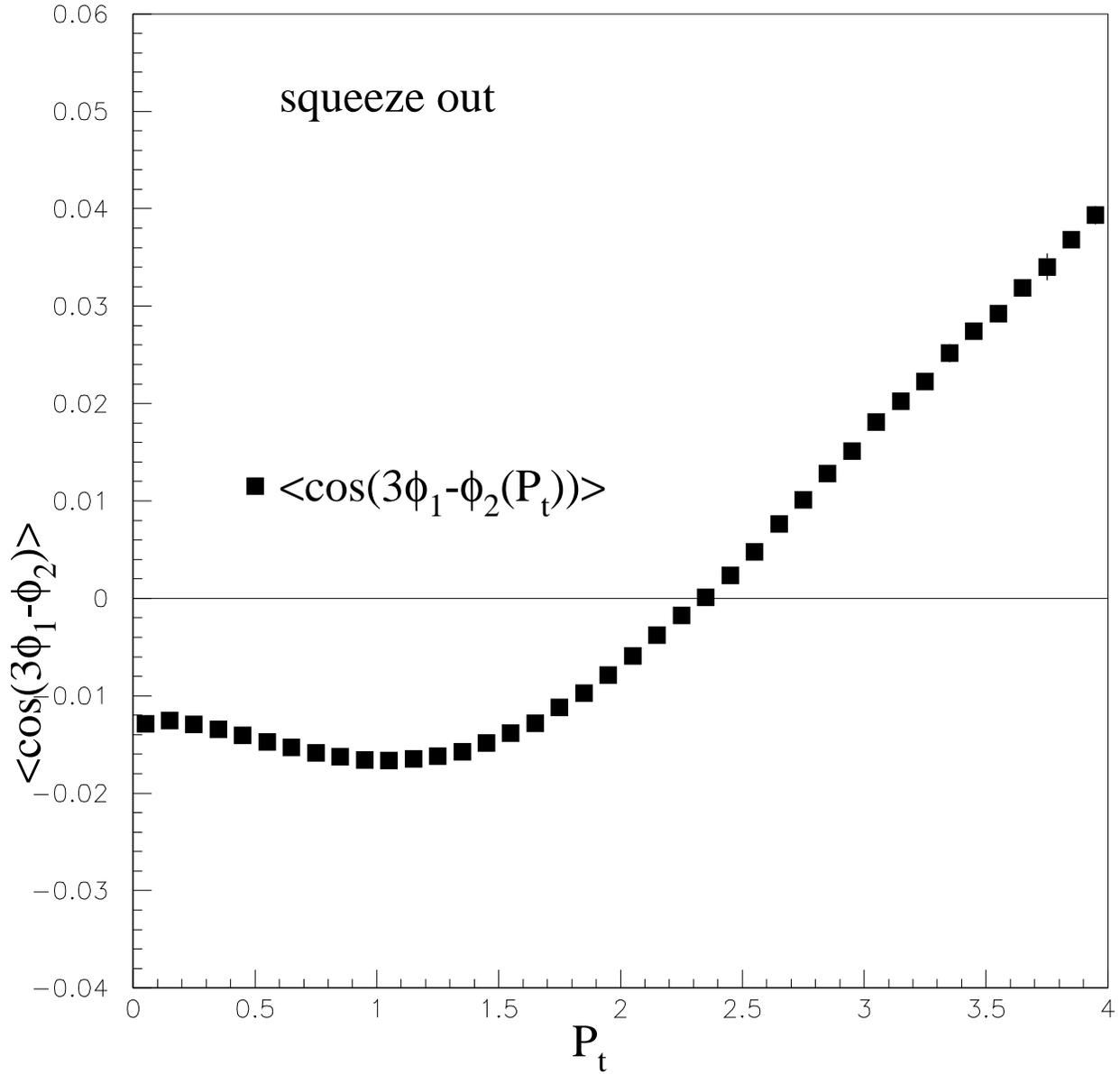}}
\end{center}
\vspace{2pt}
\caption{ An angular correlation between $v_3$ and $p_t$ dependent $v_1$ with 
respect to the $v_2$ reaction plane. $\langle cos(3\phi_1 - \phi_2(p_t) - 
2\Psi_{RP})\rangle$  is a sum over the simulated central events(see text) and 
is a two particle correlation using particle 1 and 2 since the reaction plane 
is a fixed axis constructed by the squeeze out particles.}
\label{fig2}
\end{figure}
 
\begin{figure}
\begin{center}
\mbox{
   \epsfysize 7.0in
   \epsfbox{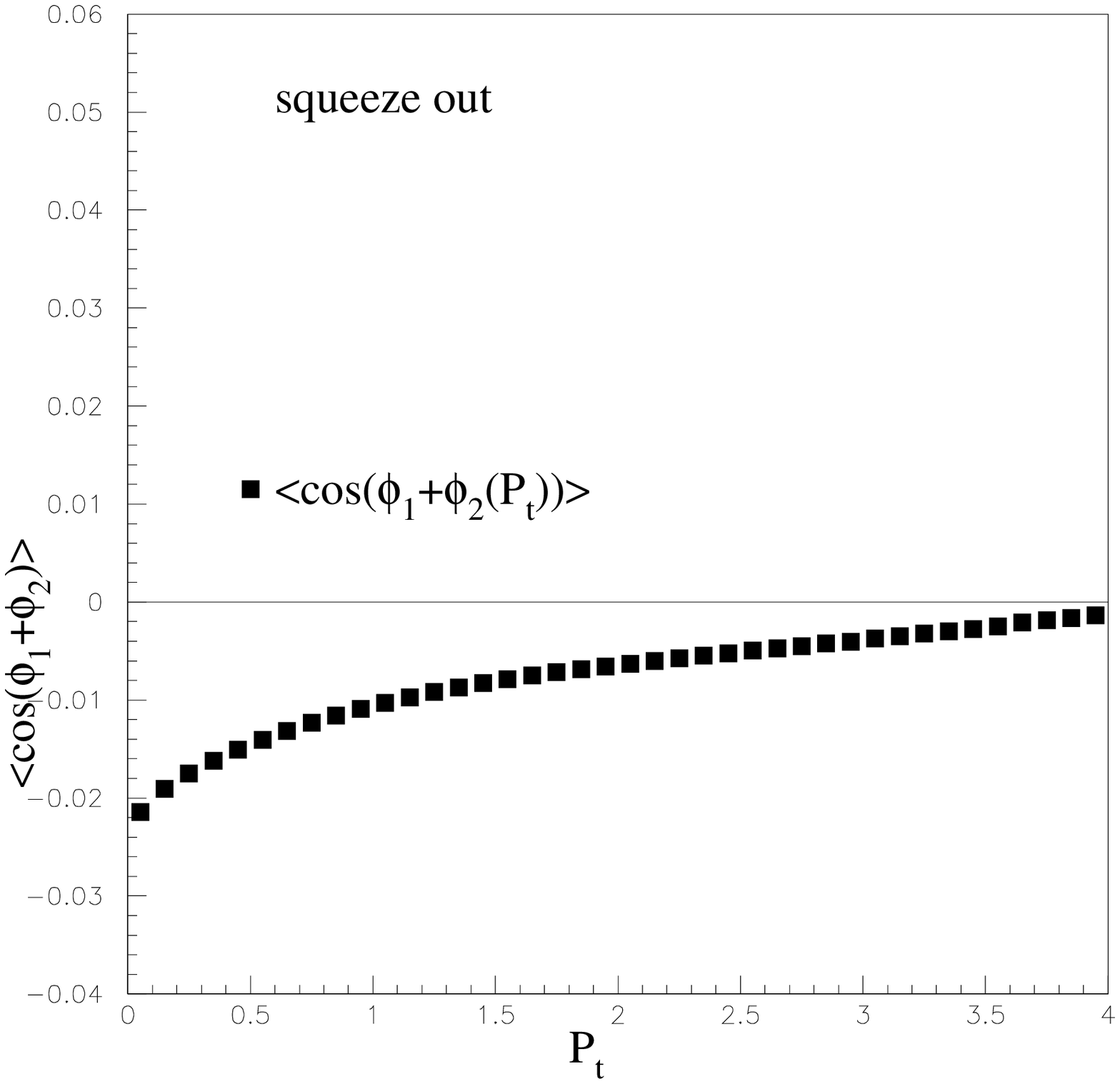}}
\end{center}
\vspace{2pt}
\caption{ An angular correlation between $v_1$ and $p_t$ dependent $v_1$ with 
respect to the $v_2$ reaction plane. $\langle cos(\phi_1 + \phi_2(p_t) - 
2\Psi_{RP})\rangle$  is a sum over the simulated central events(see text) and 
is a two particle correlation using particle 1 and 2 since the reaction plane 
is a fixed axis constructed by the squeeze out particles.}
\label{fig3}
\end{figure}

\begin{figure}
\begin{center}
\mbox{
   \epsfysize 7.0in
   \epsfbox{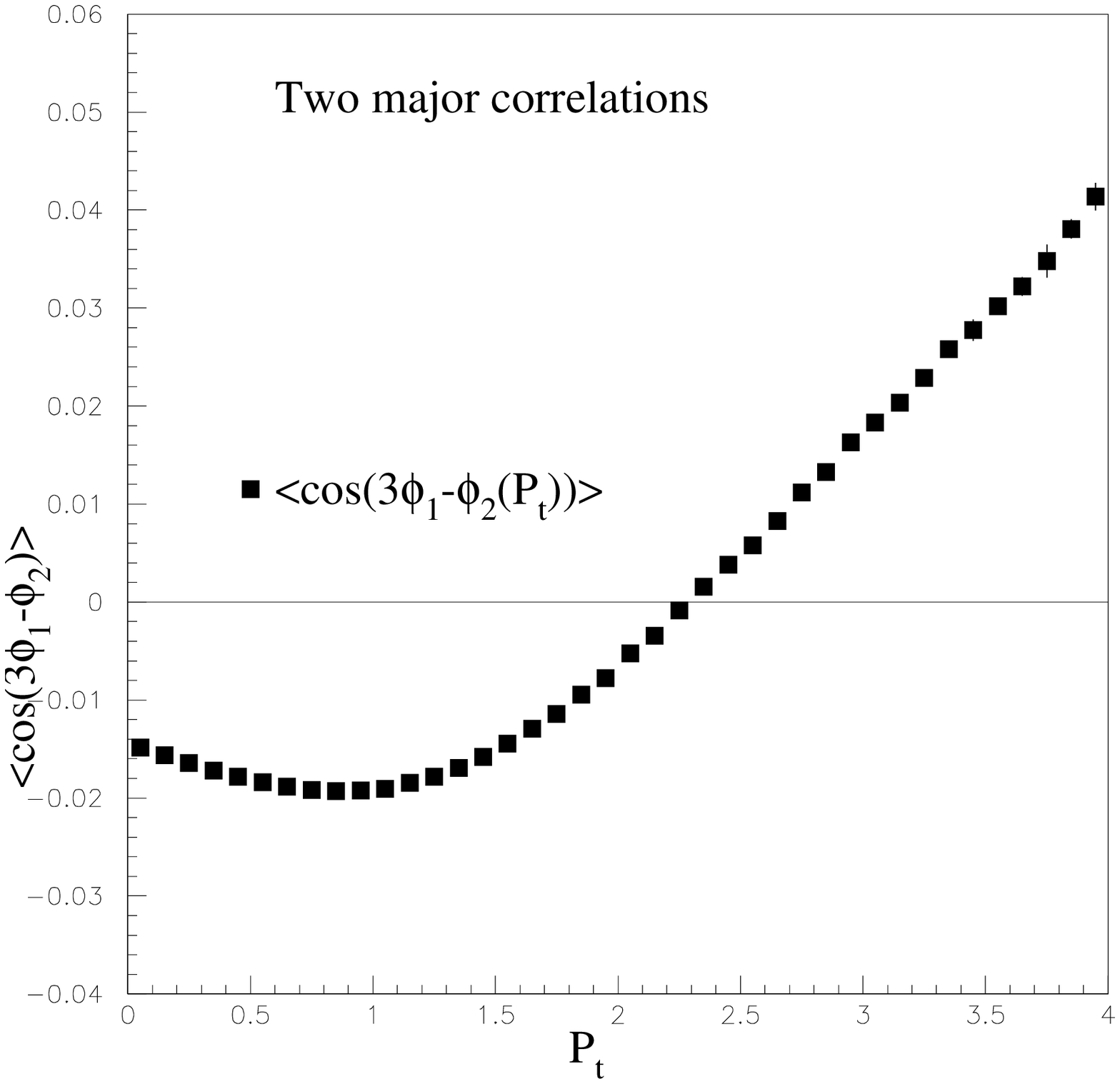}}
\end{center}
\vspace{2pt}
\caption{ An angular correlation between $v_3$ and $p_t$ dependent $v_1$ with 
respect to the $v_2$ reaction plane one particle coming from the flux tubes 
with another coming from the squeeze out particles. $\langle cos(3\phi_1 - 
\phi_2(p_t) - 2\Psi_{RP})\rangle$  is a sum over subset of particles in 
the simulated central events(see text) and is a two particle correlation using 
particle 1 and 2 since the reaction plane is a fixed axis constructed by the 
squeeze out particles. We retain the normalization of all pairs and only use 
subsets of particle pairs in the correlation.}
\label{fig4}
\end{figure}
 
\begin{figure}
\begin{center}
\mbox{
   \epsfysize 7.0in
   \epsfbox{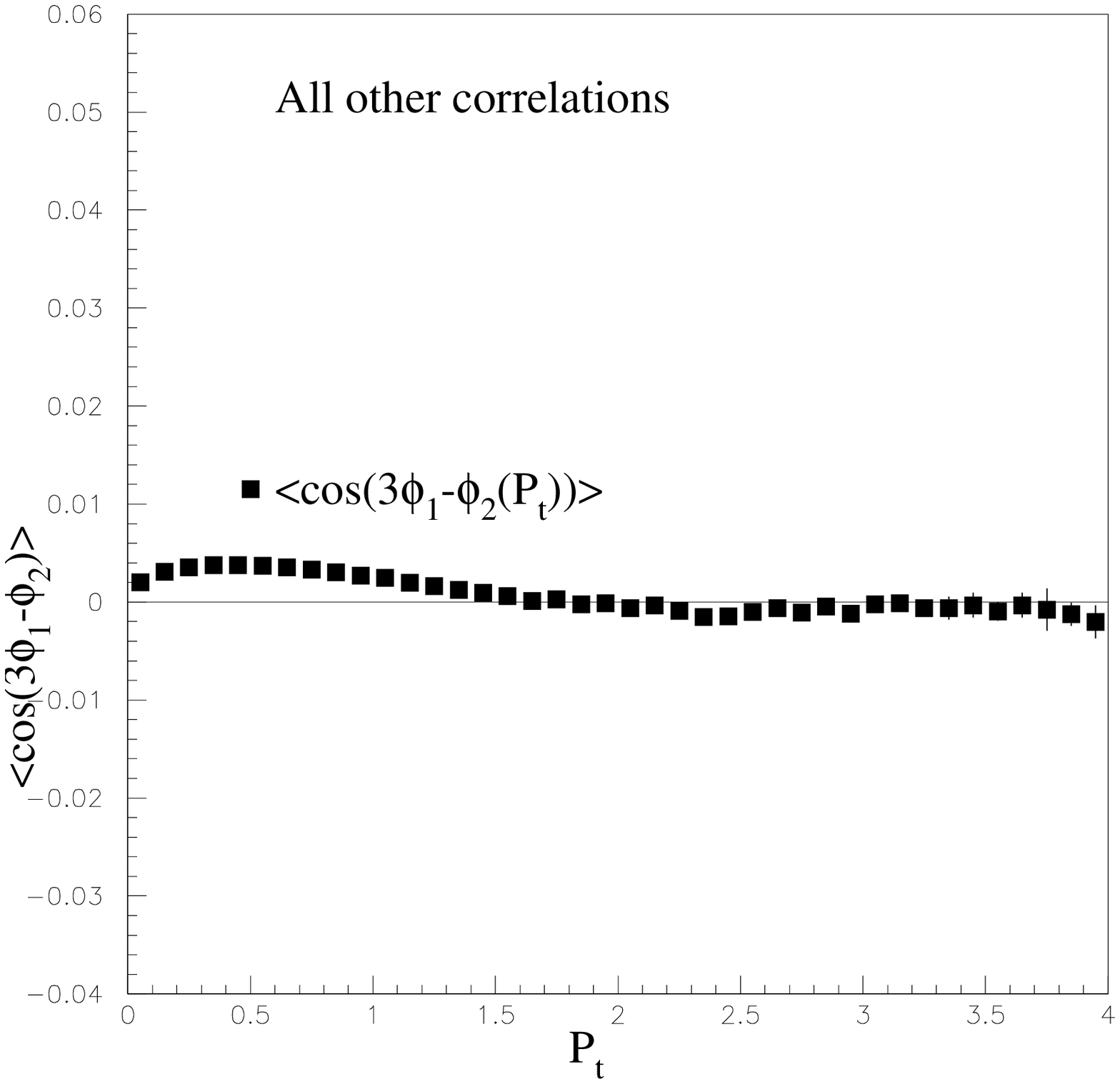}}
\end{center}
\vspace{2pt}
\caption{ An angular correlation between $v_3$ and $p_t$ dependent $v_1$ with 
respect to the $v_2$ reaction plane using particle not coming from pairs 
between the flux tubes and the squeeze out particles. $\langle cos(3\phi_1 - 
\phi_2(p_t) - 2\Psi_{RP})\rangle$  is a sum over subset of particles in the 
simulated central events(see text) and is a two particle correlation using 
particle 1 and 2 since the reaction plane is a fixed axis constructed by 
the squeeze out particles. We retain the normalization of all pairs and only 
use subsets of particle pairs in the correlation. We see that all the other 
pairs pairs except those between the flux tubes and the squeeze out particles 
seem to more or less cancel out giving us insight into the 
drivers of the observed correlation.}
\label{fig5}
\end{figure}

\section{Two Major Components to the Angular Correlation between $v_3$ and 
$p_t$ dependent $v_1$ with respect to the $v_2$ reaction plane}

In our simulation the flux tubes and the squeeze out particles have unique
tags thus making it possible to explore their contribution to the over all
correlation function show in Figure 2. Each of the $p_t$ bins of the 
correlation in Figure 2. has a normalization. If we retain this normalization
and only use subsets of particle pairs, we achieve the same correlation function
when we add up all the different pair correlations. For this section we will 
consider one particle coming from the flux tubes with another coming from the
squeeze out particles. This sub correlation function is shown in Figure 4.
 
If we consider all the other pairs of particles leaving out the above pairs
between the flux tubes and the squeeze out particles we arrive at the
correlation of Figure 5. We see that all the other pairs seem to more or less
cancel out giving us insight into the drivers of the observed correlation. 
Let us look into the flux tube part of the pair correlation. The largest flux
tube has a special tag(tag 1). In Figure 6. we plot the correlation of all
tag 1 particles(largest flux tube) with the squeeze out particles. We see that 
in this correlation the $v_3$ of the squeeze out particles are pointing in the 
$180^0$ direction the same as the global $v_3$, while the largest flux tube 
$v_1$ points in the $0^0$ direction(see Figure 1.). With increasing $p_t$ this 
correlation becomes stronger and stronger, because there is an increase in 
higher $p_t$ particles coming from the strongest flux tube.

The other flux tubes on the surface conserve the momentum of the strongest
flux tube thus generate a $v_1$ along the $180^0$ direction. This makes the
squeeze out $v_3$(same as the global $v_3$) point in the same direction 
generating a negative correlation between the particles of the other flux tubes
and the squeeze out particles. This correlation is shown in Figure 7. and is
very constant in $p_t$ because the other flux tube particles are more spread
out in $p_t$.   

\begin{figure}
\begin{center}
\mbox{
   \epsfysize 7.0in
   \epsfbox{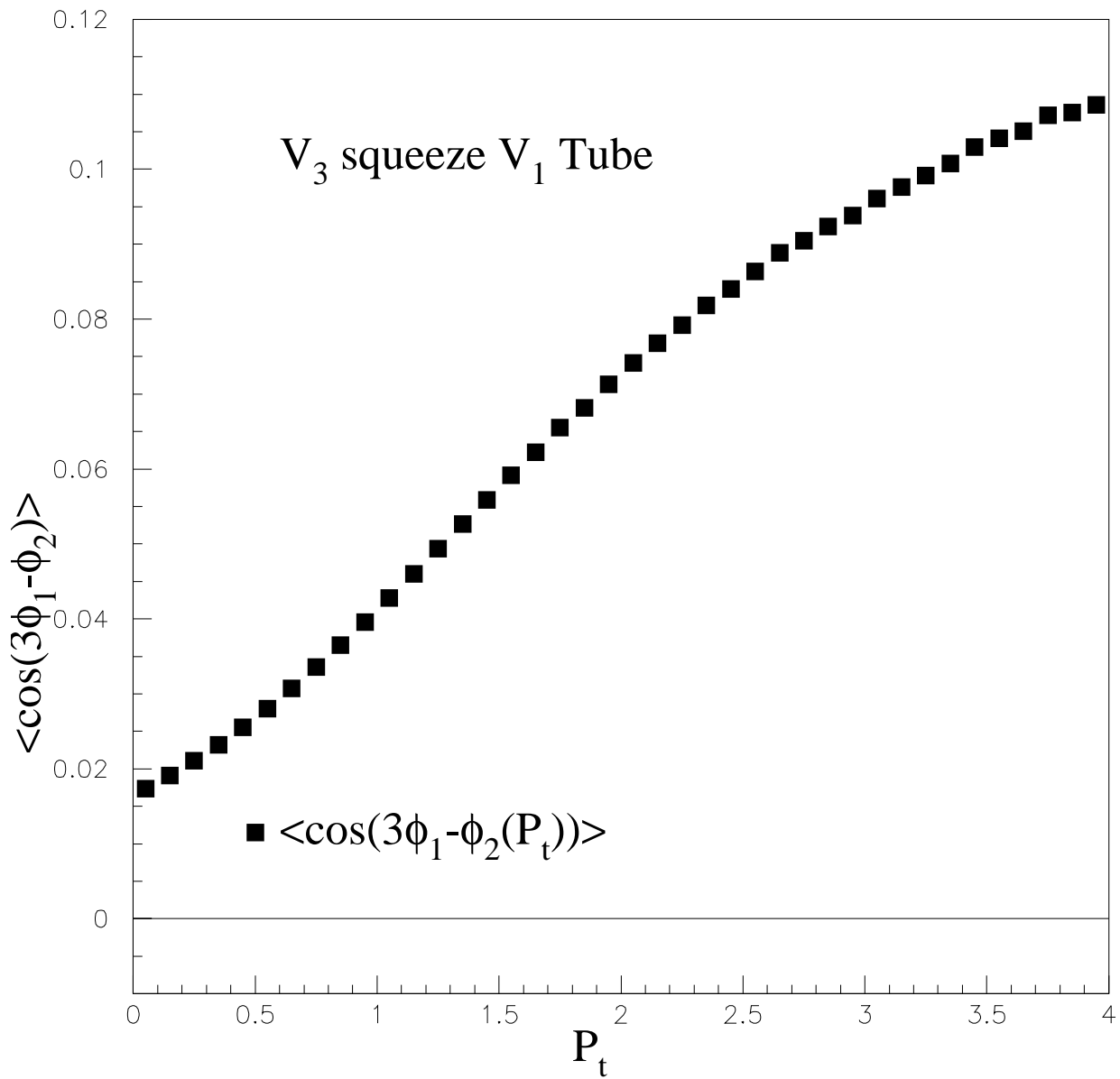}}
\end{center}
\vspace{2pt}
\caption{ An angular correlation between $v_3$ and $p_t$ dependent $v_1$ with 
respect to the $v_2$ reaction plane one particle coming from the largest flux
tube and the other coming from the squeeze out particles. $\langle cos(3\phi_1 
- \phi_2(p_t) - 2\Psi_{RP})\rangle$ is a sum over subset of particles in 
the simulated central events(see text) and is a two particle correlation using 
particle 1 and 2 since the reaction plane is a fixed axis constructed by the 
squeeze out particles. We retain the normalization of all pairs and only use 
subsets of particle pairs in the correlation. We see a large increase in a 
positive correlation with $p_t$ since the largest flux tube 
particles($v_1$) are back to back to the squeeze out particles($v_3$). This 
gives us insight into the drivers of the observed correlation of Figure 2..}
\label{fig6}
\end{figure}

\begin{figure}
\begin{center}
\mbox{
   \epsfysize 7.0in
   \epsfbox{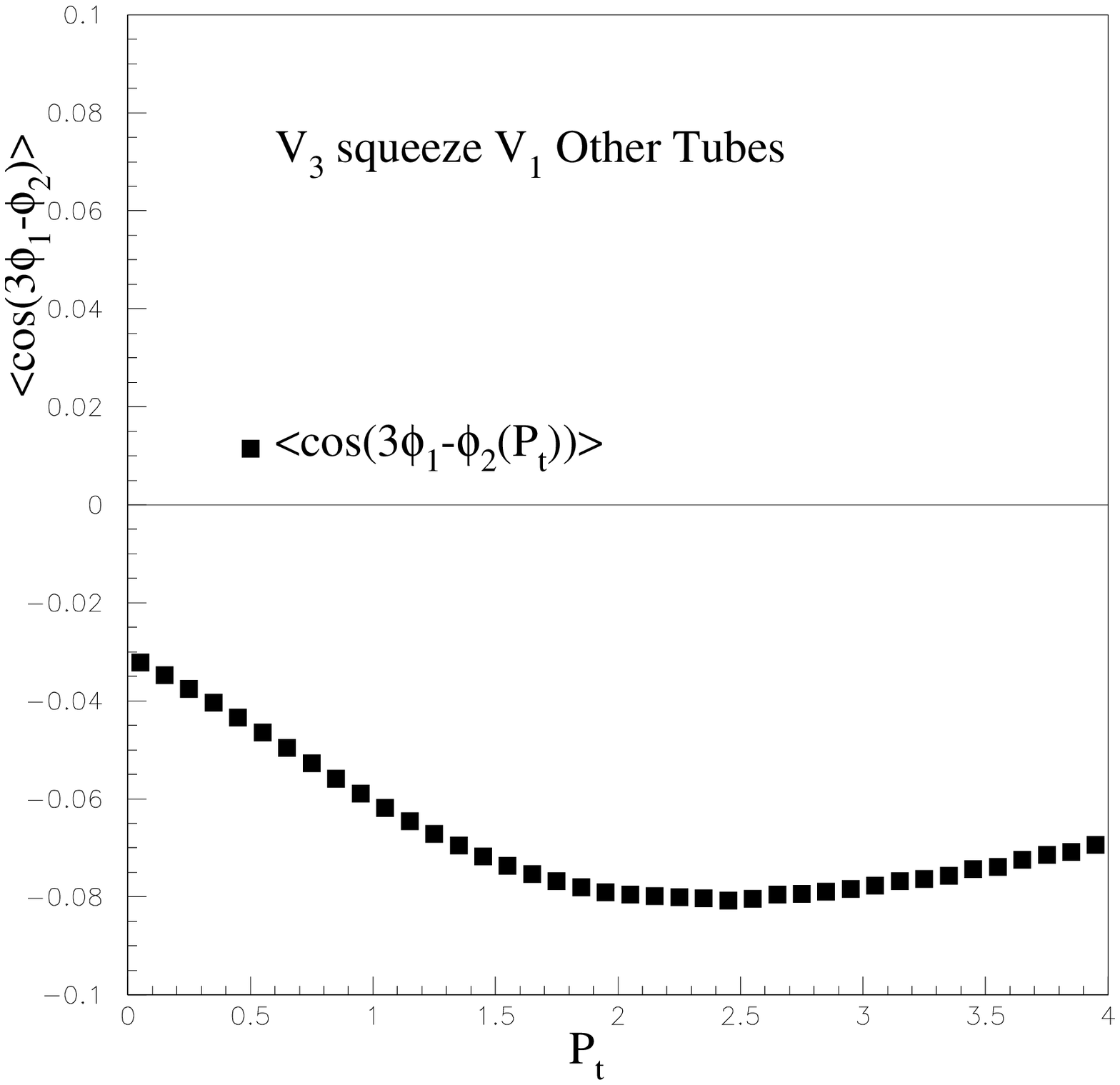}}
\end{center}
\vspace{2pt}
\caption{ An angular correlation between $v_3$ and $p_t$ dependent $v_1$ with 
respect to the $v_2$ reaction plane one particle coming from the other flux
tubes with the second coming from the squeeze out particles. 
$\langle cos(3\phi_1 - \phi_2(p_t) - 2\Psi_{RP})\rangle$ is a sum over subset 
of particles in the simulated central events(see text) and is a two particle 
correlation using particle 1 and 2 since the reaction plane is a fixed axis 
constructed by the squeeze out particles. We retain the normalization of all 
pairs and only use subsets of particle pairs in the correlation. We see that 
the correlation is negative since the $v_1$ of the other flux tubes particles
and the $v_3$  of the squeeze out particles point in the same direction. The 
correlation is very flat in $p_t$ because of wider range of values of 
particle $p_t$ in the other flux tubes.}
\label{fig7}
\end{figure}

\section{The same Two Major Components to the Angular Correlation between 
$v_1$ and $p_t$ dependent $v_1$ with respect to the $v_2$ reaction plane}

We will explore the same two major correlations of the last section since
these correlation gave us insight into the overall correlations in the
squeeze out system. We will use the same unique tags thus making it possible 
to fallow the $v_1$ in $p_t$ with respect to the global $v_1$ and explore 
their contribution to the over all correlation function show in Figure 3.
As we did in the last section we will consider one particle coming from the 
flux tubes with another coming from the squeeze out particles. This sub 
correlation function is shown in Figure 8. This correlation shows a negative 
and quit flat correlation of momentum conservation between the flux tubes and 
the squeeze out particles.

If we consider all the other pairs of particles leaving out the above pairs
between the flux tubes and the squeeze out particles we arrive at the
correlation of Figure 9. We see that all the other pairs except those between 
the flux tubes and the squeeze out particles seem to be aligned with a positive
correlation acting together to conserve momentum of the flux tube and squeeze 
out system. The largest flux tube has a special tag(tag 1). In Figure 10. we 
plot the correlation of all tag 1 particles(largest flux tube) with the 
squeeze out particles. We see that the large flux tube and the squeeze out 
particles have a negative correlation from momentum conservation.

The other flux tubes on the surface conserve the momentum of the strongest
flux tube thus generate a $v_1$ along the $180^0$ direction. This makes the
squeeze out $v_1$ which is also pointing in the $180^0$ direction be aligned
at high $p_t$ thus drives a positive correlation. However as a whole the flux 
tube system forms a momentum conserving system. The softer particles of the
other flux tubes are a part of this system and have the momentum direction of 
the system which generates a negative correlation between the squeeze out 
particles and flux tubes(see Figure 11.).

\begin{figure}
\begin{center}
\mbox{
   \epsfysize 7.0in
   \epsfbox{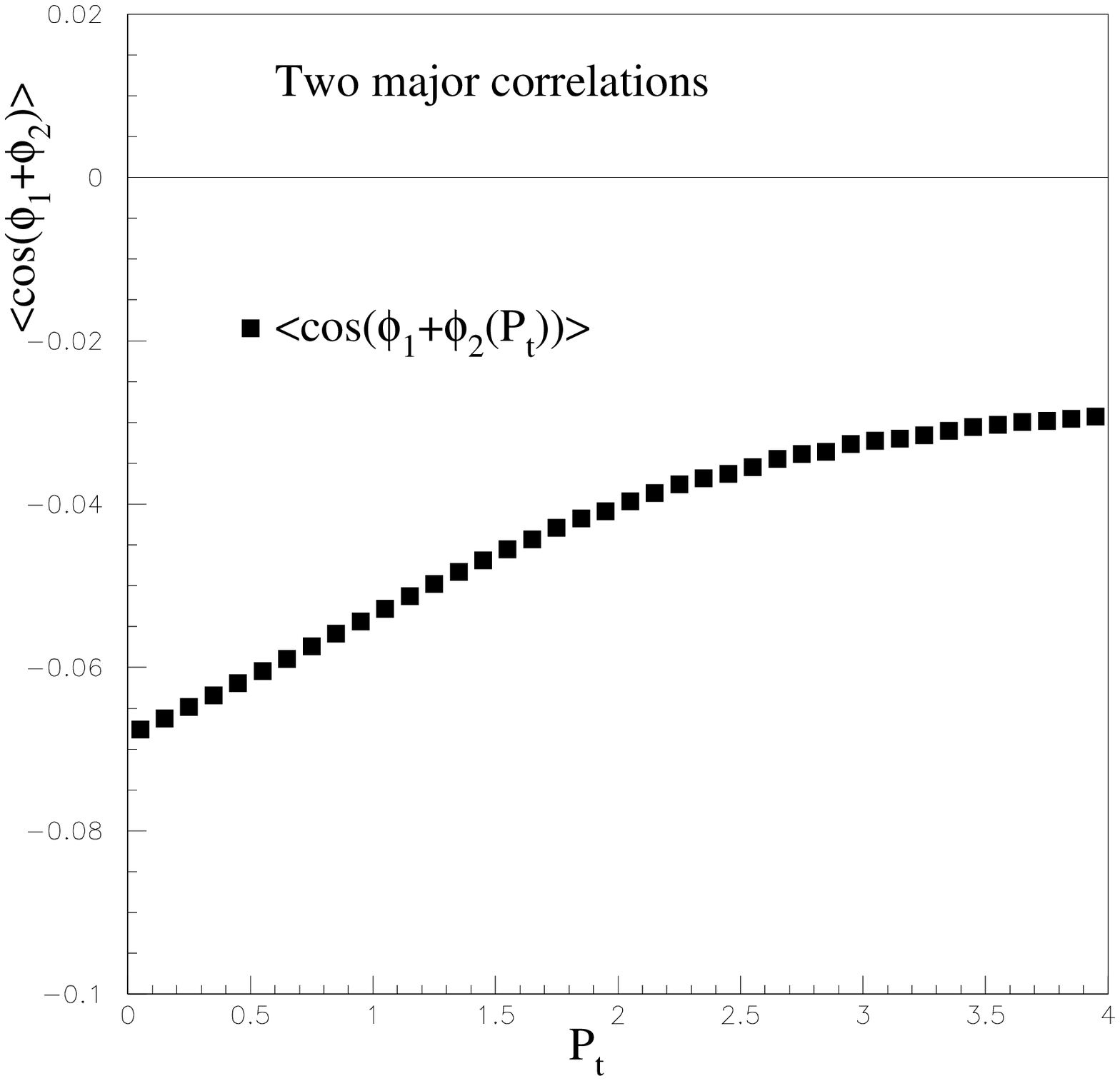}}
\end{center}
\vspace{2pt}
\caption{ An angular correlation between $v_1$ and $p_t$ dependent $v_1$ with 
respect to the $v_2$ reaction plane one particle coming from the flux
tubes and the other coming from the squeeze out particles. $\langle cos(\phi_1 
+ \phi_2(p_t) - 2\Psi_{RP})\rangle$ is a sum over subset of particles in 
the simulated central events(see text) and is a two particle correlation using 
particle 1 and 2 since the reaction plane is a fixed axis constructed by the 
squeeze out particles. We retain the normalization of all pairs and only use 
subsets of particle pairs in the correlation. We see that the correlation is
negative and quit flat showing momentum conservation between tha flux tubes 
and the squeeze out particles.}
\label{fig8}
\end{figure}

\begin{figure}
\begin{center}
\mbox{
   \epsfysize 7.0in
   \epsfbox{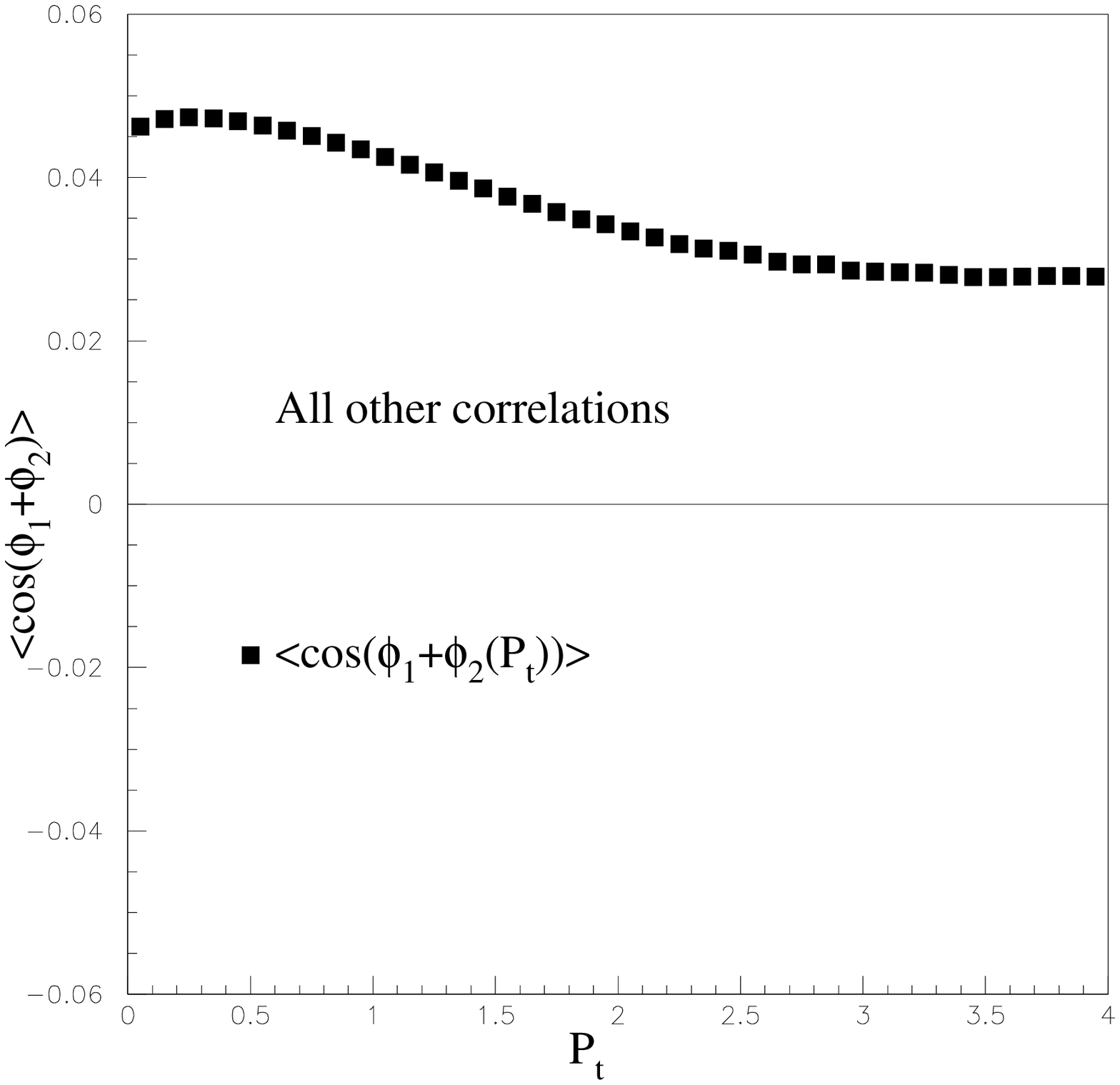}}
\end{center}
\vspace{2pt}
\caption{ An angular correlation between $v_1$ and $p_t$ dependent $v_1$ with 
respect to the $v_2$ reaction plane using particle not coming from pairs 
between the flux tubes and the squeeze out particles. $\langle cos(\phi_1 +
\phi_2(p_t) - 2\Psi_{RP})\rangle$  is a sum over subset of particles in the 
simulated central events(see text) and is a two particle correlation using 
particle 1 and 2 since the reaction plane is a fixed axis constructed by 
the squeeze out particles. We retain the normalization of all pairs and only 
use subsets of particle pairs in the correlation. We see that all the other 
pairs except those between the flux tubes and the squeeze out particles 
seem to be aligned with a positive correlation acting together to conserve
momentum of the flux tube and squeeze out system.}
\label{fig9}
\end{figure}

\begin{figure}
\begin{center}
\mbox{
   \epsfysize 7.0in
   \epsfbox{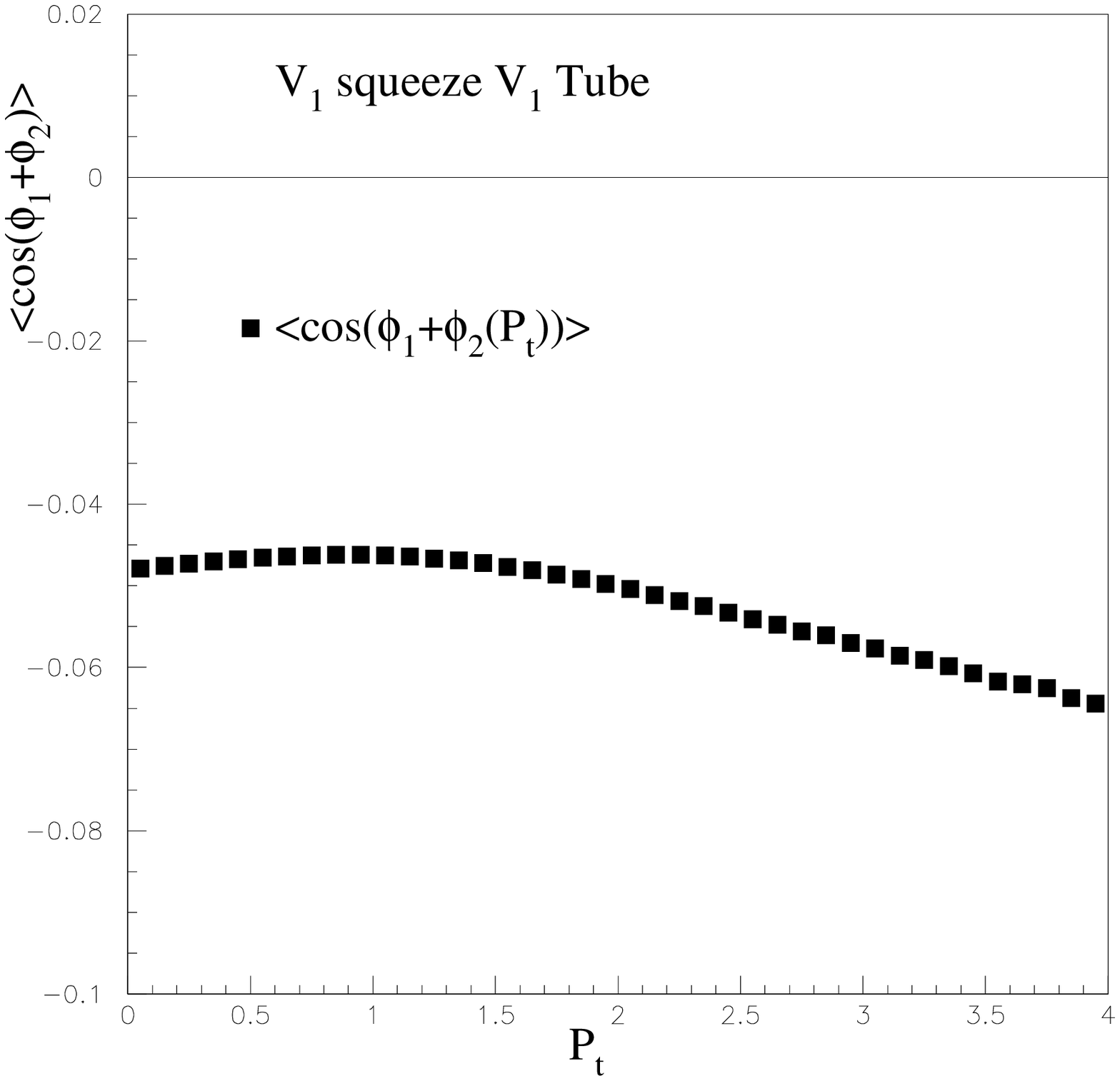}}
\end{center}
\vspace{2pt}
\caption{ An angular correlation between $v_1$ and $p_t$ dependent $v_1$ with 
respect to the $v_2$ reaction plane one particle coming from the largest flux
tube and the other coming from the squeeze out particles. $\langle cos(\phi_1 
+ \phi_2(p_t) - 2\Psi_{RP})\rangle$ is a sum over subset of particles in 
the simulated central events(see text) and is a two particle correlation using 
particle 1 and 2 since the reaction plane is a fixed axis constructed by the 
squeeze out particles. We retain the normalization of all pairs and only use 
subsets of particle pairs in the correlation. We see that the large flux tube
and the squeeze out particles have a negative correlation from momentum
conservation.}
\label{fig10}
\end{figure}

\begin{figure}
\begin{center}
\mbox{
   \epsfysize 7.0in
   \epsfbox{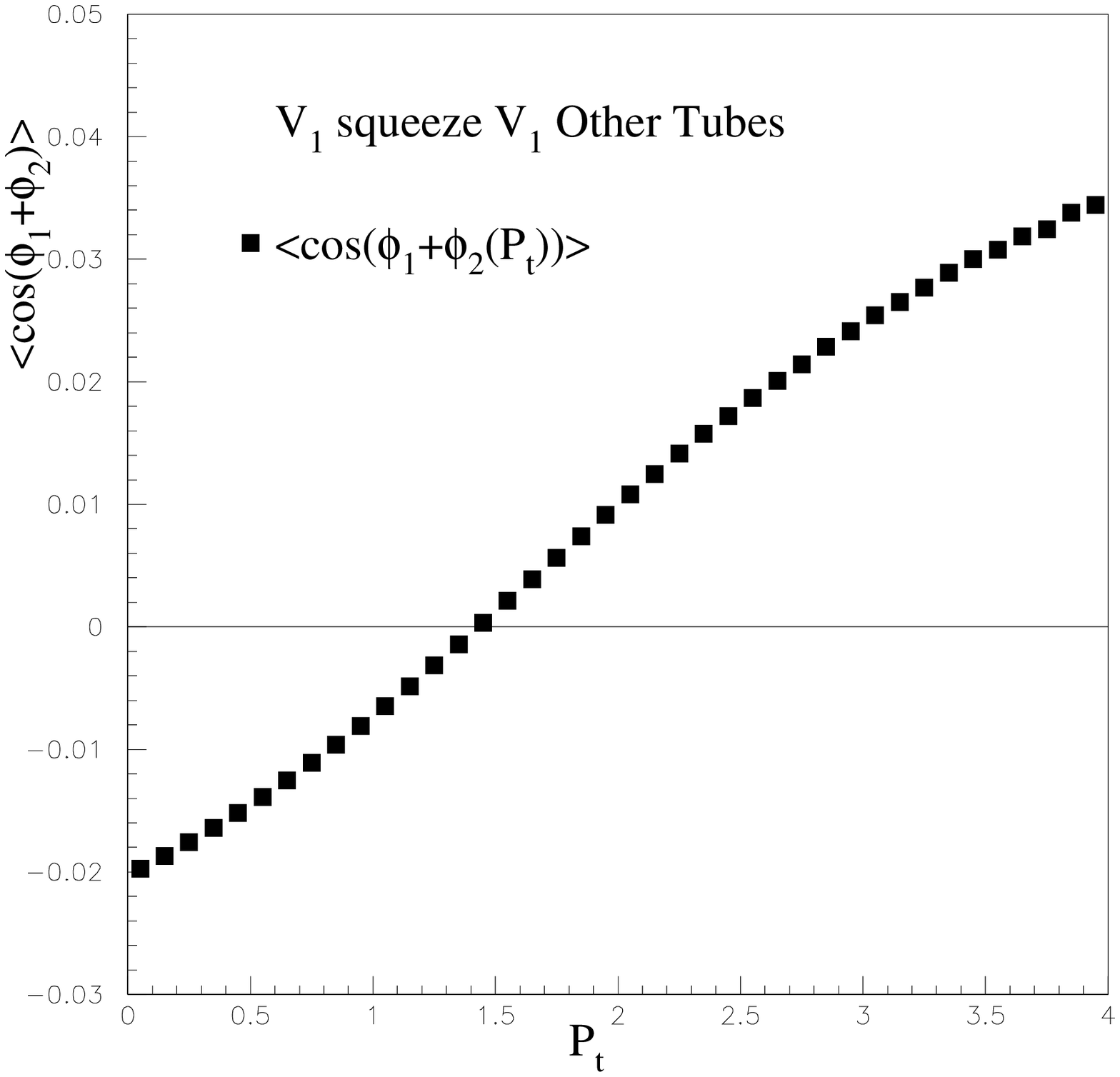}}
\end{center}
\vspace{2pt}
\caption{ An angular correlation between $v1$ and $p_t$ dependent $v_1$ with 
respect to the $v_2$ reaction plane one particle coming from the other flux
tubes with the second coming from the squeeze out particles. 
$\langle cos(\phi_1 + \phi_2(p_t) - 2\Psi_{RP})\rangle$ is a sum over subset 
of particles in the simulated central events(see text) and is a two particle 
correlation using particle 1 and 2 since the reaction plane is a fixed axis 
constructed by the squeeze out particles. We retain the normalization of all 
pairs and only use subsets of particle pairs in the correlation. We see that 
at low $p_t$ there is a negative correlation between the particles of the
flux tubes and the particles of the squeeze out(conservation of momentum 
between flux tube system). At higher $p_t$ the momentum conserving between 
the largest flux tube and the other flux tubes then align the other flux 
tube particles with the squeeze out particles and drives a positive 
correlation.}
\label{fig11}
\end{figure}

\section{Summary and Discussion}
The flux tube model does well in describing a central Au-Au collision at 
RHIC\cite{QGP,tubevsjet}. However the hydro flow of particles quarks and gluons
around the largest flux tube on the fireball surface which we will call 
squeeze out is predicted \cite{shadowing}. This is an analogy to tooth paste 
squeezing out the sides of a tooth paste tube opening when the opening is 
blocked by a dried and harden tooth paste plug. This squeeze out has been 
simulated in Ref.\cite{shadowing} but was called flux tube shadowing. 
The major conclusion of flux tube shadowing from the above reference is the 
development of a strong $v_3$ azimuthal correlation. This $v_3$ will extend 
over the length of the flux tube which was created by the initial glasma 
conditions\cite{QGP}. We needed to add to the flux tube model of 
Ref.\cite{tubevsjet} a hydro squeeze out around the largest flux tube. By 
hand we will add this modification and explore possible two particle 
correlations with respect the flow axis($v_2$) defined by the squeeze out 
particles. This flow axis will be at right angles the the largest flux tube.

We have created a  simulation of squeeze out to have a global $v_3$ pointing
in the minus X direction(see lower figure of Figure 1.). At low $p_t$ $v_1$ 
will also point in this minus X direction. However as we move to higher $p_t$ 
the largest flux tube will take over and $v_1$ will switch to the positive X 
axis. We can capture this behavior through a two particle
correlation  with respect to the reaction plane($\Psi_{RP}$) given by global 
$v_2$.

\begin{equation}
\langle cos(3\phi_1 - \phi_2(p_t) - 2\Psi_{RP})\rangle ,
\end{equation}
where $\Psi_{RP}$, $\phi_1$, $\phi_2$ denote the azimuthal angles of the
reaction plane, produced particle 1, and produced particle 2. This two
particle azimuthal correlation measures the difference between the global $v_3$
and the $p_t$ dependent $v_1$. If we would rotate all events such that 
$\Psi_{RP}$ = 0.0, then we have
\begin{equation}
\langle cos(3\phi_1 - \phi_2(p_t))\rangle.
\end{equation}

We calculate this correlation over the simulated central events and obtain the
values shown in Figure 2. where the $p_t$ is that of the second particle.
This sign change moving from lower $p_t$ to higher $p_t$ is the shift of $v_1$
from aligned with global $v_3$ at lower $p_t$ to anti-aligned at higher $p_t$.

In our simulation the flux tubes and the squeeze out particles have unique
tags thus making it possible to explore their contribution to the over all
correlation function show in Figure 2.  We can achieve about the same 
correlation as Figure 2 if we consider particle coming from the flux tubes 
with particles coming from the squeeze out particles(Figure 4.). 

The largest flux tube has a special tag. In Figure 6. we plot the correlation 
of the largest flux tube particles($v_1$) and particles of the squeeze(global 
$v_3$).  The largest flux tube $v_1$ points in the $0^0$ 
direction(see Figure 1.) while squeeze out particles($v_3$) point in the 
$180^0$(see Figure 1.). With increasing $p_t$ this correlation becomes 
stronger and stronger, because there is an increase in higher $p_t$ particles
coming from the strongest flux tube.

The other flux tubes on the surface conserve the momentum of the strongest
flux tube thus generate a $v_1$ along the $180^0$ direction. This makes the
squeeze out $v_3$(same as the global $v_3$) point in the same direction 
generating a negative correlation between the particles of the other flux tubes
and the squeeze out particles. This correlation is shown in Figure 7. and is
very constant in $p_t$ because the other flux tube particles are more spread
out in $p_t$.   

This squeeze out particle flow is a true hydro effect and can be detected
using the correlation functions presented in this write up. In RHIC
central Au-Au collision there is a very large $v_2$,$v_3$ and $p_t$ dependent
$v_1$ which is driven by this largest flux tube with squeeze out flowing 
particles around the tube.

\section{Acknowledgments}

This research was supported by the U.S. Department of Energy under Contract No.
DE-AC02-98CH10886.

\end{document}